\documentclass[aps,prb,preprint,groupedaddress,showpacs,amsmath,amssymb]{revtex4}

\usepackage{graphicx}
\usepackage{dcolumn}
\usepackage{bm}

\begin{document}

\title{Magnetic field tuning of the low temperature state in YbNiSi$_3$}

\author{Sergey L. Bud'ko, Paul C. Canfield}
\affiliation{Ames Laboratory US DOE and Department of Physics and Astronomy, Iowa State University, Ames, IA
50011, USA}
\author{Marcos A. Avila, Toshiro Takabatake}
\affiliation{Department of Quantum Matter, ADSM, Hiroshima University, Higashi-Hiroshima, Hiroshima 739-8530,
Japan}

\date{\today}

\begin{abstract}
We present detailed, low temperature, magnetoresistance and specific heat data of single crystal YbNiSi$_3$
measured in magnetic field applied along the easy magnetic axis, $H \| b$. An initially antiferromagnetic ground
state changes into a field-induced metamagnetic phase at $\sim 16$ kOe ($T \to 0$). On further increase of
magnetic field, magnetic order is suppressed at $\sim 85$ kOe. The functional behaviors of the resistivity and
specific heat are discussed in comparison with those of the few other stoichiometric, heavy fermion compounds with
established field-induced quantum critical point.
\end{abstract}



\maketitle

The "quantum criticality conundrum" \cite{lau01a} continues to be a focus of attention for many theorists and
experimentalists. \cite{ste01a,ste06a,col01a,sac00a,geg03a,con05a,col05a} In the particular case of magnetic
intermetallic compounds that manifest heavy fermion ground states, magnetic ordering temperatures can be tuned by
some control parameter to $T = 0$, bringing the material to a quantum critical point (QCP). This tuning is
possible because of a delicate balance between the Ruderman - Kittel - Kasuya - Yoshida (RKKY) interaction, that
defines the magnetic ordering temperature, and the Kondo effect, that causes screening of the magnetic moments and
favors non-magnetic ground state.\cite{don77b} Recently, in addition to traditionally used pressure and chemical
substitution, magnetic field has also emerged as a potential control parameter,\cite{ste01a,geg03a} having an
advantage of being a continuous and common parameter for a number of experimental techniques. However, the
applicability  of a single theoretical description for a QCP reached by using different tuning parameters, is
still under discussion.

So far the number of {\it stoichiometric} materials exhibiting a field-induced QCP is rather small, with only few
of them being Yb-based: YbRh$_2$Si$_2$,\cite{geg02a} YbAgGe,\cite{bud04a} YbPtIn,\cite{mor06a}. The release of
entropy associated with the magnetic ordering (which can be used as a rough caliper of the sise of the Yb magnetic
moment in the ordered state) in the three materials ranges from $\sim 0.01 R\ln 2$ for YbRh$_2$Si$_2$, to $\leq
0.1 R\ln 2$ for YbAgGe, to $\sim 0.6 R\ln 2$ for YbPtIn. Nevertheless the $H - T$ phase diagrams and functional
behavior of a number of physical parameters in the vicinity of the field-induced QCP are very similar for all
three of these compounds. Recently investigated YbNiSi$_3$,\cite{avi04a} an orthorhombic, moderately -  heavy
fermion compound with the N\'eel temperature, $T_N = 5.1$ K, Sommerfeld coefficient, $\gamma \approx 190$ mJ/mol
K$^2$, and entropy associated with the magnetic ordering $\sim 0.6 R\ln 2$, seems to be a suitable candidate for
further studies of the possible field-induced quantum criticality in stoichiometric, Yb-based, heavy fermions.

Single crystals of YbNiSi$_3$ were grown from Sn flux (see Ref. \onlinecite{avi04a} for more details). Flux
residue from the surface of the samples was polished or/and etched out to exclude the effect of elemental Sn on
measured electrical transport properties. DC magnetization was measured up to 55 kOe and down to 1.8 K in a
Quantum Design MPMS-5 SQUID magnetometer. Heat capacity and standard, 4-probe, AC ($f = 16$ Hz) resistivity in
zero and applied field up to 140 kOe were measured in a Quantum Design PPMS-14 instrument using a $^3$He
refrigerator with heat capacity and ACT options respectively. In all measurements an external field was applied
along easy magnetic axis ($H \| b$).\cite{avi04a} Resistivity in applied field was measured in a transverse,  $H
\perp I$, configuration.

\begin{figure}
\begin{center}
\includegraphics[angle=0,width=120mm]{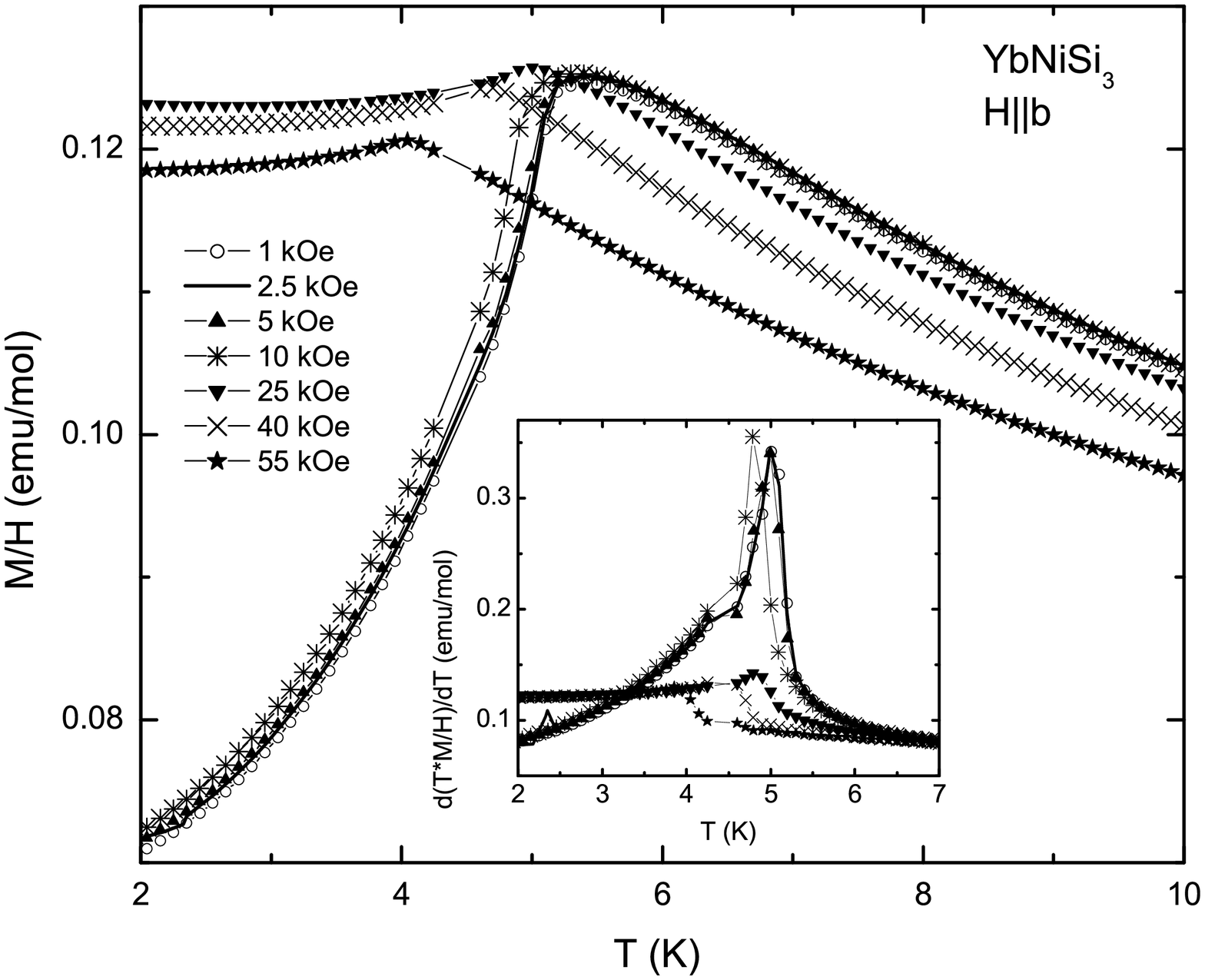}
\end{center}
\caption{Low temperature DC magnetic susceptibility, $M(T)/H$, measured in YbNiSi$_3$ in different applied fields,
$H \| b$. Inset: $d(\chi T)/dT$ corresponding to the data in the main panel. Gap in the data near between 4.3 K
and 4.6 K corresponds to the region of unstable temperature control in our MPMS-5 magnetometer.}\label{F0}
\end{figure}

\begin{figure}
\begin{center}
\includegraphics[angle=0,width=120mm]{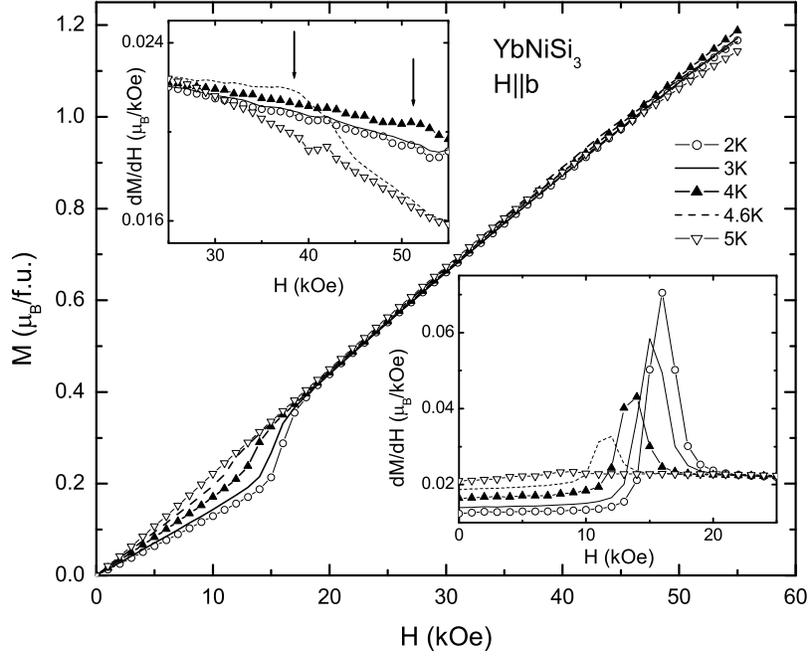}
\end{center}
\caption{Representative magnetic isotherms, $M(H)$, measured in YbNiSi$_3$ at different temperatures for $H \| b$.
Insets: enlarged low- and high-field plots of corresponding $dM/dH$. Arrows on upper left inset mark second, high
field, feature, that broadens beyond resolution for $T = 5$ K.}\label{F1}
\end{figure}

The low field magnetic susceptibility, $\chi = M/H$, measurements and 2 K magnetization, $M(H)$, isotherms are
consistent with the data reported in Ref. \onlinecite{avi04a}. The feature associated with the antiferromagnetic
transition in $\chi$ (or in $d(\chi T)/dT$)\cite{fis62a} shifts to lower temperatures with increase of the
magnetic field applied (Fig. \ref{F0}). The lower field metamagnetic transition (Fig. \ref{F1}) moves to lower
fields with increase of temperature at which the $M(H)$ data were taken, at higher temperatures the second feature
moves to accessible field range (upper left inset to Fig. \ref{F1}); this feature corresponds to high field break
in $M(H)$ slope reported previously \cite{avi04a}.

\begin{figure}
\begin{center}
\includegraphics[angle=0,width=120mm]{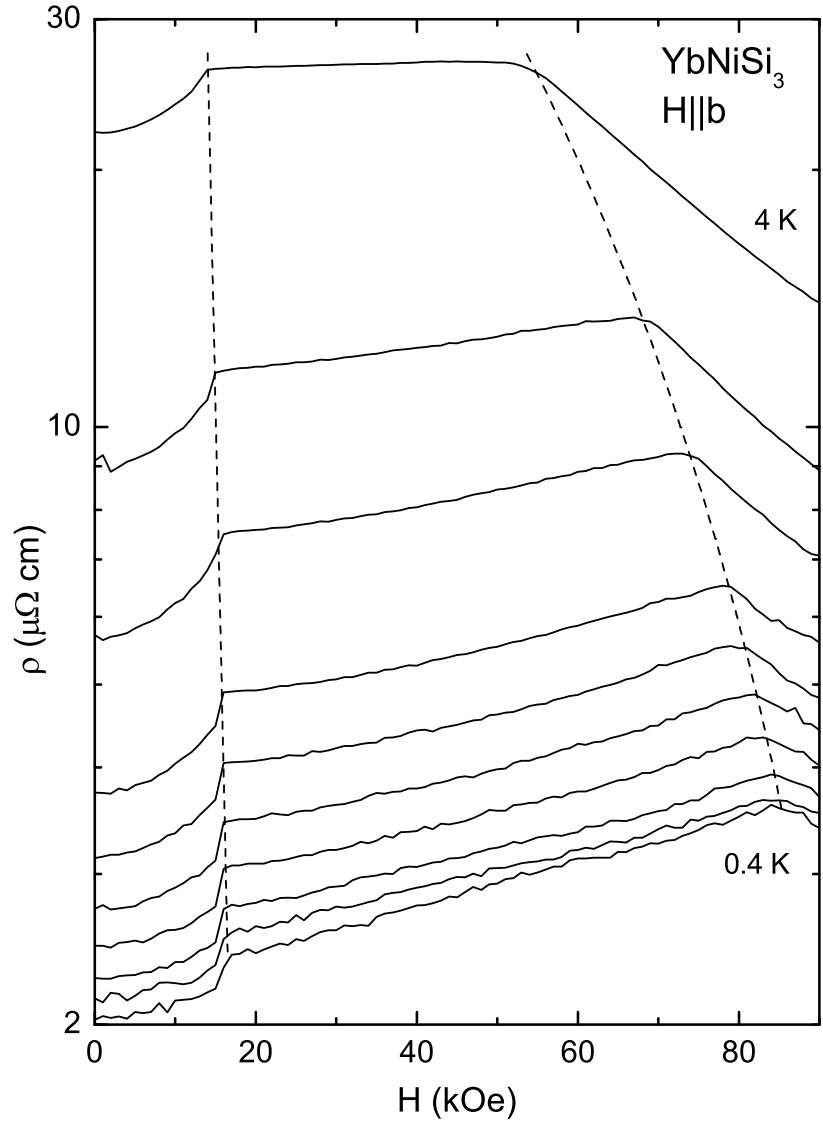}
\end{center}
\caption{Low temperature resistivity isotherms, $\rho(H)$, measured at (from the bottom to the top) 0.4, 0.76, 1,
1.25, 1.5, 1.75, 2, 2.5, 3, and 4 K. Dashed lines are guide for the eye.}\label{F2}
\end{figure}

Field-dependent resistivity isotherms are shown in Fig. \ref{F2}. Two features (marked with dashed lines),
consistent with those seen in aforementioned magnetization ($M(H)$) measurements, are easily recognizable and can
be monitored in the $H - T$ domain of the experiments. The position of the lower field feature is almost
temperature-independent, while the critical field associated with the higher field feature decreases with an
increase of temperature.

\begin{figure}
\begin{center}
\includegraphics[angle=0,width=120mm]{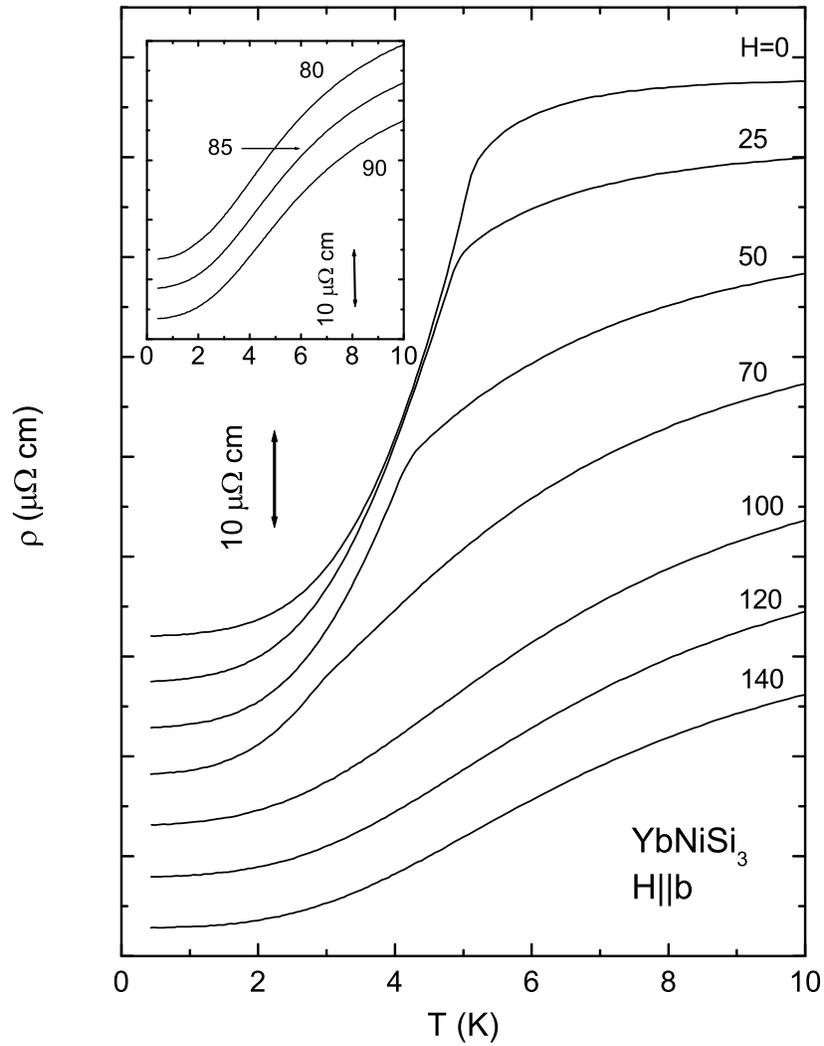}
\end{center}
\caption{Representative low temperature resistivity curves measured in zero and applied magnetic field. Curves are
shifted down along the $Y$-axis by multiplicative of 5 $\mu\Omega$ cm for clarity. Numbers correspond to applied
field in kOe.}\label{F3}
\end{figure}

Representative-temperature dependent resistivity, $\rho(T)$, curves are shown in Fig \ref{F3}. No traces of the
signal from residual Sn flux are seen in $H = 0$ data. The resistivity at base temperature, $T \approx 400$ mK, in
zero field is $ \approx 2 \mu\Omega$ cm, consistent with $\rho_0 = 1.5 \mu\Omega$ cm obtained from $\rho = \rho_0
+ AT^2$ fit in [\onlinecite{avi04a}]. A clear break of the $\rho(T)$ slope and the lower temperature decrease of
resistivity below the magnetic ordering temperature, $T_N$, associated with a loss of spin-disorder scattering is
observed up to 70 kOe. At higher fields no signature of magnetic transition can be seen. Moreover, there in no
clear signature of traditional form of non-Fermi-liquid behavior (linear or close-to-linear temperature dependence
of resistivity) at fields close to the one at which $T_N$ is suppressed to $T = 0$ (Fig. \ref{F3}, inset).

\begin{figure}
\begin{center}
\includegraphics[angle=0,width=90mm]{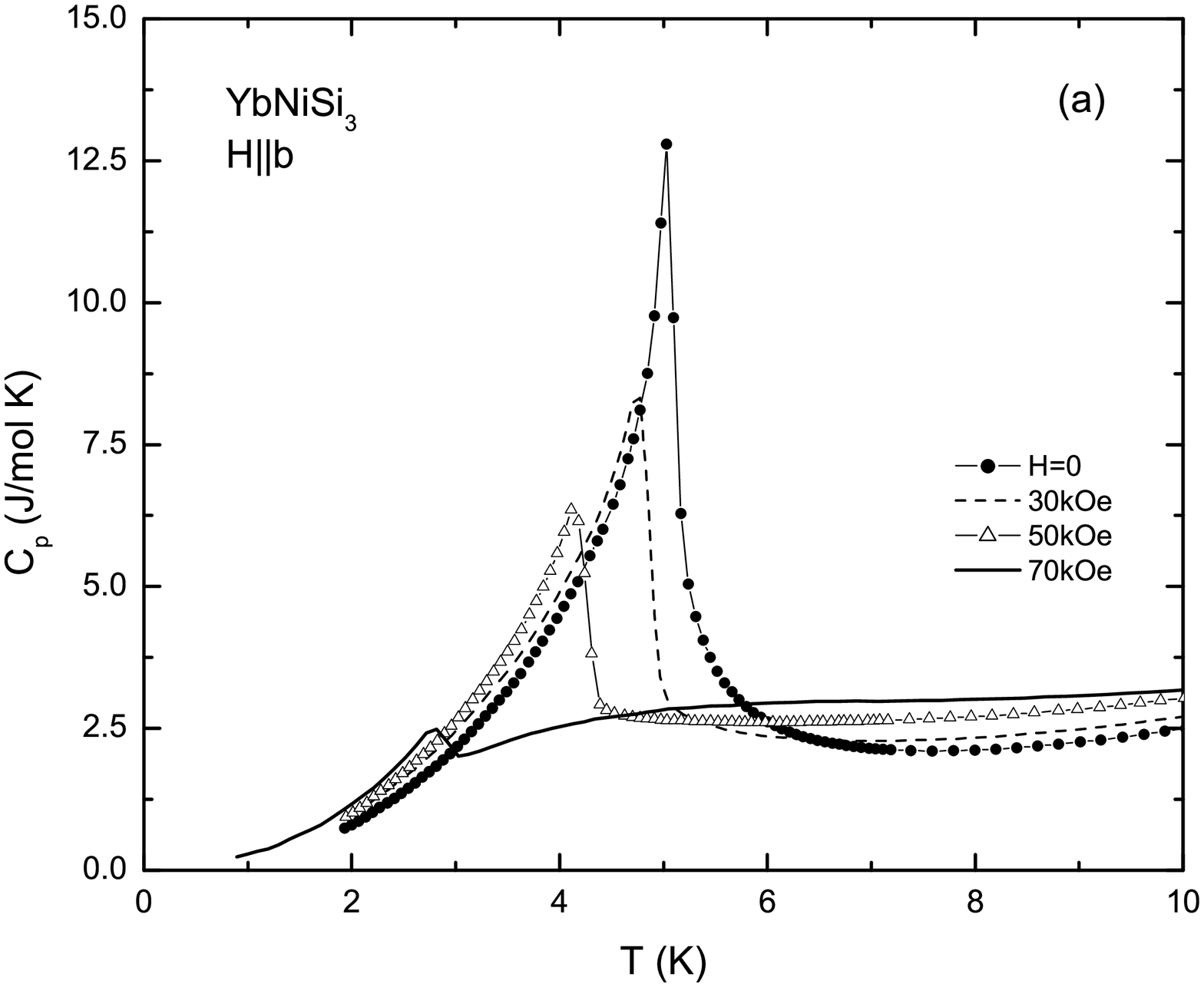}
\includegraphics[angle=0,width=90mm]{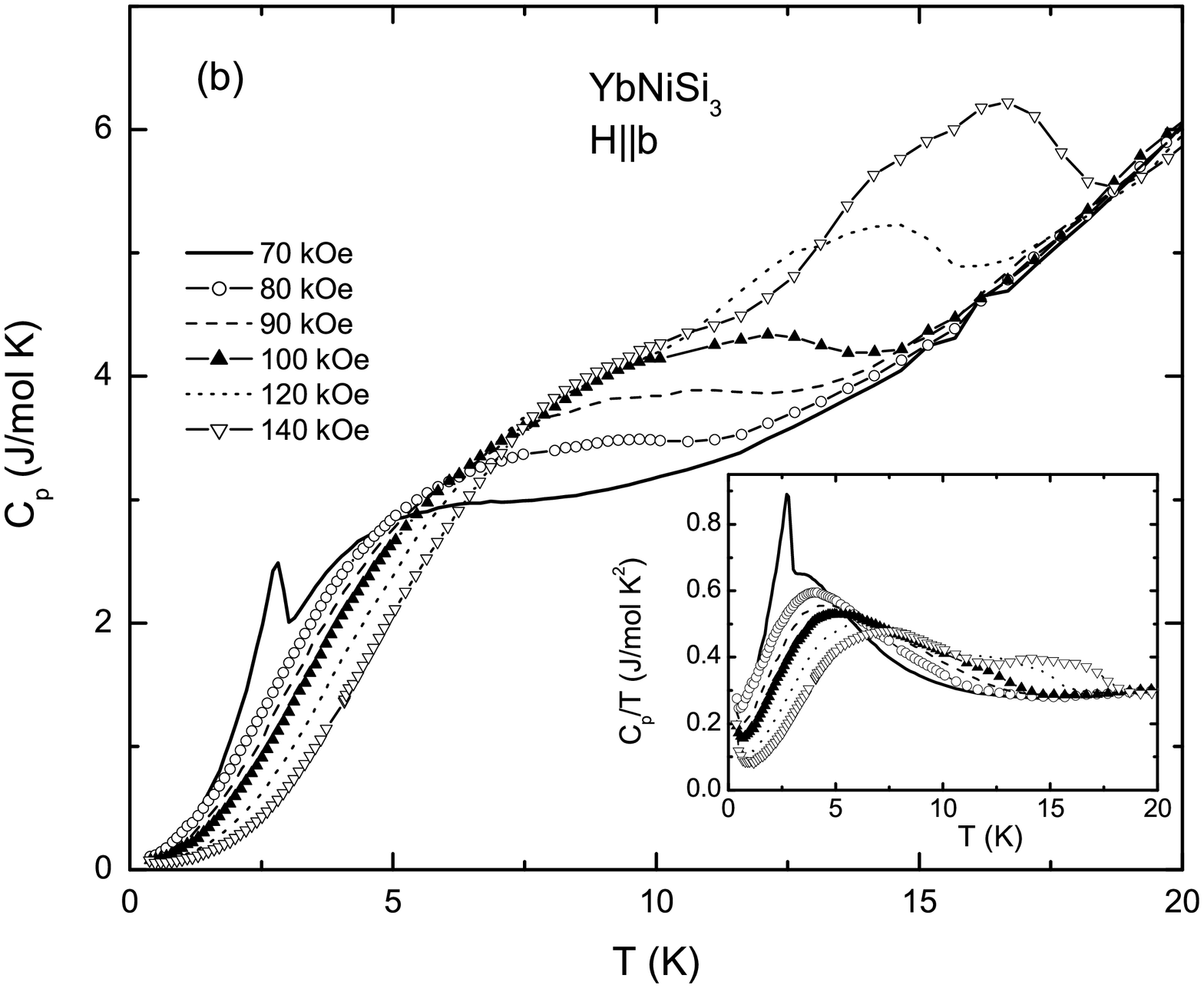}
\end{center}
\caption{Low temperature heat capacity, $C_p(T)$, measured in different applied magnetic fields. Inset to panel
(b): $C_p/T$ as a function of temperature. Note that the curve for $H = 70$ kOe is shown on both
panels.}\label{F4}
\end{figure}

The temperature-dependent heat capacity is shown in Fig. \ref{F4}. Results in zero field are consistent with the
published data.\cite{avi04a} Consistent to aforesaid, the peak in $C_p(T)$ moves down with an increase of the
applied magnetic field, it's height decreases, and it is not recognized in the data  at 80 kOe and above. Closer
examination of $H = 70$ kOe curve (Fig. \ref{F4}b) reveals that the peak associated with the magnetic long range
order is located on a background with a broad maximum. It is seen with more clarity if $C_p/T$ is plotted as a
function of temperature (inset to Fig. \ref{F4}b). This broad maximum moves up in temperature and acquires
structure with an increase of applied field: the data for $H = 120$ and 140 kOe can be deconvoluted to up to three
broad peaks. For $H \leq 50$ kOe these broad non-monotonic background is probably obscured by the feature in
$C_p(T)$ associated with the magnetic transition. The plausible explanation for these broad features is that they
are Schottky-like contributions to the specific heat: the the degeneracy of crystal-electric-field-defined lower
energy levels of YbNiSi$_3$ (e.g. a quadruplet or two closely spaced doublets) is lifted by the Zeeman term of the
hamiltonian, and the energy difference between different levels increases further on increase of the applied
field. This explanation is in in dispute with the assumption of the doublet ground state of YbNiSi$_3$
\cite{avi04a} and calls for additional studies.

\begin{figure}
\begin{center}
\includegraphics[angle=0,width=120mm]{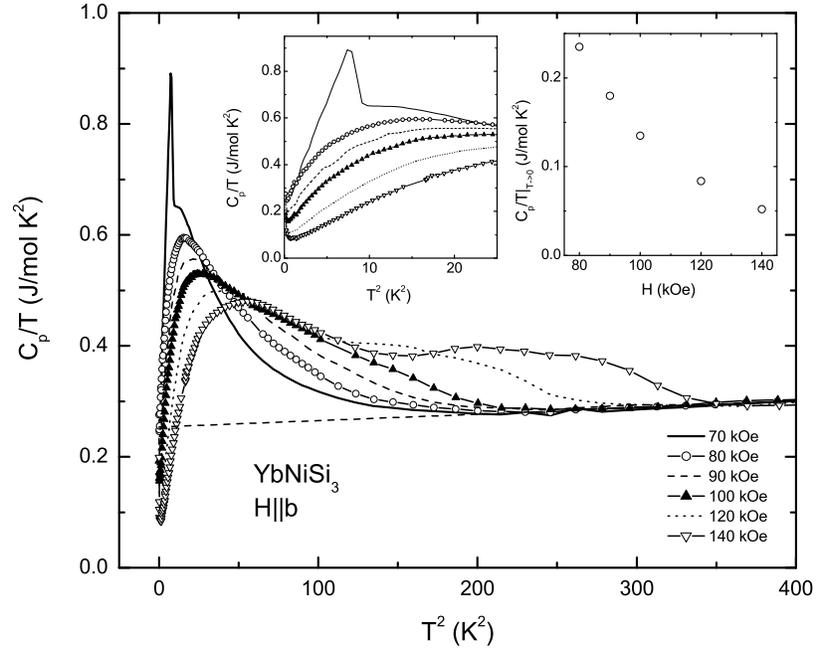}
\end{center}
\caption{Heat capacity in applied field for 70 kOe $\leq H \leq$ 140 kOe plotted as $C_p/T$ vs. $T^2$. Dashed line
- linear extrapolation from the high temperature data. Insets: left - enlarged low temperature part; right -
estimate of $\gamma$ from extrapolated low temperature linear portion of the $C_p/T$ vs. $T^2$ graph.}\label{F7}
\end{figure}

In order to evaluate the change of the electronic specific heat coefficient, $\gamma$, heat capacity data are
re-plotted as $C_p/T$ vs. $T^2$ in Fig. \ref{F7}. The small upturn at low temperatures (left inset to Fig.
\ref{F7}) that becomes more pronounced at higher fields is probably associated with the nuclear Schottky
contribution to the specific heat. The Sommerfeld coefficient in the specific heat, $\gamma$, can be estimated by
extrapolating the high temperature linear part of the $C_p/T$ vs. $T^2$ curves to $T = 0$ (dashed line in Fig.
\ref{F7}). As a result of such procedure $\gamma \approx 250$ mJ/mol K$^2$, practically independent of the applied
field can be assessed. This value is close to the estimate in Ref. \onlinecite{avi04a} for $H = 0$. However, this
estimate exceeds significantly the values of $C_p/T$ measured in $T \to 0$ regime for magnetic fields in which the
magnetic order is suppressed, pointing out that, at least for 70 kOe $\leq H \leq$ 140 kOe, high temperature
extrapolation of $C_p/T$ vs. $T^2$ is not an unique estimate of $\gamma$. This brings back the question of the
estimate of the entropy associated with the magnetic transition \cite{avi04a} and may point out that (in lieu of
the data on non-magnetic analogue) the non-magnetic contribution, taken as extrapolation from above 10 K, was
overestimated and the cited value of $\Delta S_{mag} \sim 0.6R\ln2$ was underestimated.

Operationally, one can take the extrapolation to $T \to 0$ of the lowest temperature measurements plotted as
$C_p/T$ vs. $T^2$, see left inset to Fig. \ref{F7} (ignoring the nuclear Schottky contribution), as an estimate of
$\gamma$. The obtained values (right inset to Fig. \ref{F7}) seem to follow the trend generally observed in
materials close to QCP \cite{geg02a,bud04a,mor06a,cus03a,bal05a}. However, at least in the particular case of
YbNiSi$_3$, one cannot discard the possibility that the shifts of the Schottky-like peaks to higher temperatures
with an increase of the applied field leads to a decreasing contribution (since it is taken further from the
maximum) to the apparent $\gamma = C_p/T|_{T \to 0}$, and genuine behavior of the Sommerfeld coefficient is masked
by this fastly changing contribution.
\\

\begin{figure}
\begin{center}
\includegraphics[angle=0,width=120mm]{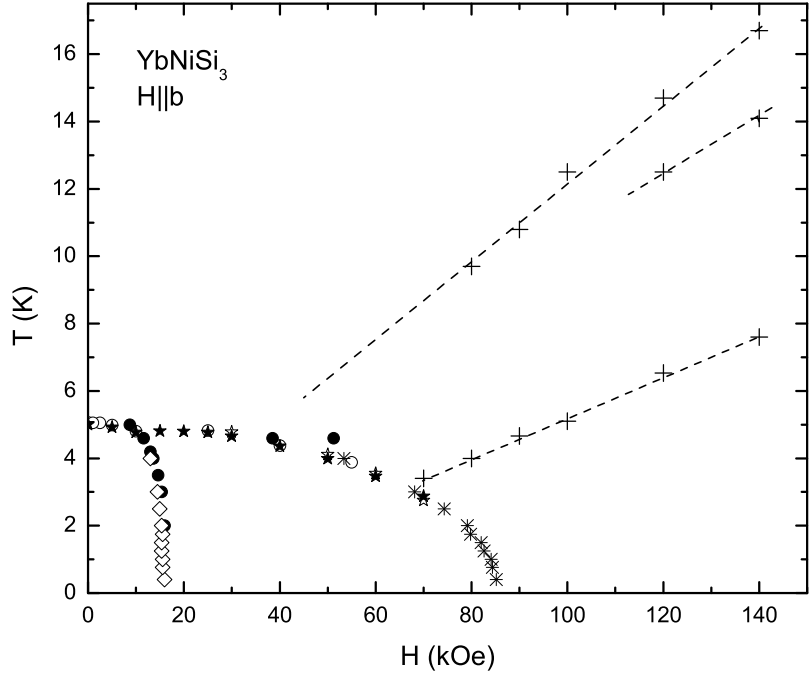}
\end{center}
\caption{Tentative $H - T$ phase diagram of YbNiSi$_3$ for $H |\ b$. Different symbols correspond to the phase
lines obtained from different thermodynamic and transport measurements: open circle - $M(T)$, filled circle -
$M(H)$, open star - $C_p(T)$, filled star - $\rho(T)$, diamond and asterisk - $\rho(H)$. Crosses correspond to the
position of broad maxima in $C_p/T$, dashed lines are guides for the eye.}\label{F5}
\end{figure}

The thermodynamic and transport measurements on YbNiSi$_3$ in magnetic field applied along the $b$-axis define the
$H - T$ phase diagram (Fig. \ref{F5}). This phase diagram is consistent (in overlapping $H - T$ domains) with the
one recently suggested \cite{gru06a} from magnetoresistance isotherms for $T \geq 1.85$ K. The phase diagram is
rather simple, with only two phases in the ordered state, and a triple point. The positions of the Schottky-like
anomalies are also shown, depending on the applied field in a close to linear fashion. A magnetic field of $\sim
85$ kOe suppresses the long range order, making the phase diagram look similarly to the cases of field-induced QCP
recently discussed in several materials. \cite{geg02a,bud04a,mor06a,cus03a,bal05a} Further understanding of the
low temperature state of YbNiSi$_3$ in applied field can be approached by detailed analysis of the
magnetotransport and thermodynamic data.

\begin{figure}
\begin{center}
\includegraphics[angle=0,width=120mm]{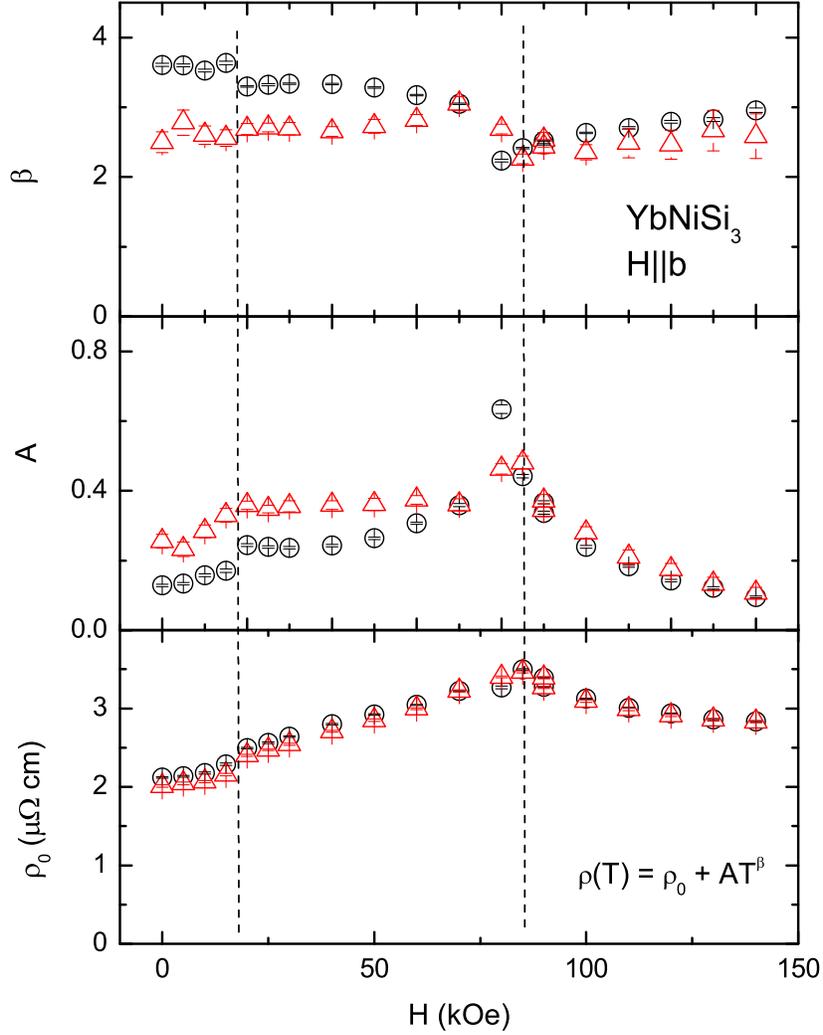}
\end{center}
\caption{Results of fitting of low temperature resistivity to the form $\rho(T) = \rho_0 + AT^\beta$; units of
$\rho(T)$ and $\rho_0$ are $\mu\Omega$ cm. Circles: fit from 3 K to the base temperature, triangles: fit from 1.5
K to the base temperature. {\it Nominal} error bars of the fits are shown. Vertical dashed lines mark $T \to 0$
critical fields from the phase diagram above.}\label{F6}
\end{figure}

Low temperature resistivity ($\rho(T)$) data measured in magnetic field were fitted with the equation $\rho(T) =
\rho_0 + AT^\beta$ (as e.g. in Ref. \onlinecite{nik06a}) with three fitting parameters: residual resistivity,
$\rho_0$, pre-factor, $A$, and exponent, $\beta$. Two sets of fits were performed: from 3 K down to base
temperature ($\approx 0.4$ K) and from 1.5 K down to base temperature. Both sets of fits are giving similar
results (Fig. \ref{F6}). All three fitting parameters have features associated with the critical fields, the lower
critical field being essentially temperature-independent below 3 K and the upper one only shifting slightly as $T$
decreases from 3 K to 1.5 K (as can be seen in $A$ and $\beta$). It is noteworthy that the exponent $\beta$ as
calculated from the data above $\approx 0.4$ K, does not fall below $\beta = 2$, and the changes in $\beta$ as
obtained from the 0.4 K $\geq T \geq$ 1.5 K are rather small (Fig. \ref{F6}, upper panel). This is clearly
different from the case of field-induced QCP in YbRh$_2$Si$_2$, \cite{geg02a,cus03a} YbAgGe \cite{bud04a,nik06a}
and other materials, although it is possible that $\beta$ approaches 1.0 for $T < 0.4$ K, and in a very narrow
field range.
\\

In summary, thermodynamic and transport measurements on YbNiSi$_3$ ($H \| b$) reveal $H - T$ phase diagram with
two distinct, antiferromagnetic and metamagnetic, ordered phases and a triple point. An applied magnetic field of
$\sim 85$ kOe suppresses the magnetic ordering temperature. As distinct from several known heavy fermions with
field induced QCP, above 400 mK no non-Fermi-signature in resistivity was observed in the vicinity of the critical
field for $T \geq 0.4$ K, although they may well emerge for $T < 0.4$ K data. Heat capacity measurements suggest
that applied magnetic field splits degenerate ground state of YbNiSi$_3$ and this field-dependent split of the
energy levels manifest itself as Schottky-like anomalies. Additionally, possible artifacts in assessing of the
Sommerfeld coefficient from heat capacity measurements, that have relevance beyond this particular compound, were
disclosed.

\begin{acknowledgments}
Ames Laboratory is operated for the U.S. Department of Energy by Iowa State University under Contract No.
W-7405-Eng.-82. Work at Ames Laboratory was supported by the Director for Energy Research, Office of Basic Energy
Sciences. Work in Hiroshima University was supported by a Grant-in-Aid for Scientific Research (COE Research
13CE2002) of MEXT Japan. PCC acknowledges R. A. Ribeiro for having devised the three-finger method of small sample
manipulation.  SLB acknowledges help of Kamotsuru on initial stage of this work.
\end{acknowledgments}

\clearpage

\clearpage

\clearpage

\clearpage

\clearpage

\clearpage

\clearpage

\clearpage

\end{document}